\documentclass[apjl]{emulateapj}
\usepackage{amsmath}
\usepackage{color}

\newcommand{\project}[1]{\textsl{#1}}
\newcommand{\jwst}{\project{JWST}}
\newcommand{\kepler}{\project{Kepler}}

\shortauthors{Kreidberg \& Loeb}

\begin{document}

\title{Prospects for Characterizing the Atmosphere of Proxima Centauri \lowercase{b}}

\author{
Laura Kreidberg\altaffilmark{1,2} \& Abraham Loeb\altaffilmark{2}
}

\email{E-mail: laura.kreidberg@cfa.harvard.edu}

\altaffiltext{1}{Harvard Society of Fellows, Harvard University, 78 Mt. Auburn St., Cambridge, MA 02138, USA}
\altaffiltext{2}{Harvard-Smithsonian Center for Astrophysics, 60 Garden Street, Cambridge, MA 02138, USA}

\begin{abstract}
The newly detected Earth-mass planet in the habitable zone of Proxima Centauri could potentially host life  -- if it has an atmosphere that supports surface liquid water. We show that thermal phase curve observations with the \project{James Webb Space Telescope} (\jwst) from $5 - 12\;\mu$m can be used to test the existence of such an atmosphere. We predict the thermal variation for a bare rock versus a planet with 35\% heat redistribution to the nightside and show that a \jwst\ phase curve measurement can distinguish between these cases at $4\,\sigma$ confidence, assuming photon-limited precision.  We also consider the case of an Earth-like atmosphere, and find that the ozone $9.8\;\mu$m band could be detected with longer integration times (a few months). We conclude that \jwst\ observations have the potential to put the first constraints on the possibility of life around the nearest star to the Solar System.
\end{abstract}


\section{Introduction}
\cite{anglada16} recently announced the exciting discovery of a potentially habitable planet orbiting our nearest neighboring star, Proxima Centauri. The planet has a minimum mass of $1.3\,M_\oplus$ and an insolation equal to two thirds that of Earth, suggesting that it could have a rocky surface with temperatures appropriate for the existence of liquid water.

Recent transit surveys such as \kepler\ have shown that Earth-like planets like these are very common --  they are found around 10 - 25\% of stars \citep[e.g.][]{petigura13, dressing13, dressing15b}.  However, none of the Earth analogs detected to date have been feasible targets for atmosphere characterization because their host stars are too distant.  At a distance of just one parsec, Proxima b provides the first opportunity for detailed characterization of a habitable world beyond the Solar System. 
An immediate question is whether Proxima b has an atmosphere at all. Tidally locked planets orbiting M-dwarfs face unique challenges to their atmospheric stability. The atmosphere may ``collapse" if the volatile inventory freezes out and becomes trapped on the nightside \citep{joshi97}. The atmosphere is also subject to erosion by stellar winds, which are denser and faster for M-dwarfs than Sun-like stars \citep{zendejas10}. Proxima also has a high rate of flaring activity that may further threaten the planet's atmosphere \citep{davenport16}. To reveal Proxima b's evolutionary history and potential for hosting life, a first step is to ascertain whether its atmosphere has survived.

\section{Possible approaches for atmosphere characterization}
\label{sec:possibilities}
Proxima b is not likely to transit its host star. The transit probability is only 1\%, and photometric transit searches have not revealed the planet \citep{kipping16}. Therefore, the methods available for characterizing the planet's atmosphere are to (i) directly image the planet, (ii) measure variations in reflected starlight with orbital phase, and (iii) measure variation in thermal emission with orbital phase.

Direct imaging is a challenge due to the small angular separation between the planet and its host star (50 mas), which is roughly an order of magnitude smaller than what has been achieved so far \citep[e.g.][]{macintosh15}. Reaching this angular separation requires very large telescope diameter: the diffraction limit of a 30 m telescope is 10 mas at $1\,\mu$m. Next-generation ground-based extremely large telescopes (ELTs) will be capable of such measurements, but they will not be available until the mid-2020s.

Method (ii) is a more promising approach for near future characterization of the planet. Variation in reflected starlight can be detected by combining high-resolution ground-based optical spectroscopy with high contrast imaging \citep{riaud07, snellen15}. The spectroscopy is sensitive to the Doppler shift of starlight reflected by the planet as it orbits, and the high contrast imaging suppresses the stellar signal.  This measurement can reveal the planet's orbital inclination and the albedo scaled by the projected area. \cite{lovis16} recently performed a feasibility study for applying this technique to Proxima b with existing observing facilities. They propose combining the SPHERE high-contrast imager and the ESPRESSO spectrograph on the Very Large Telescope (VLT), and find that the planet can be detected at $5\,\sigma$ confidence in 20 - 40 nights of observing time (assuming an Earth-like albedo). Lower albedo (expected for an airless planet) will make the detection more challenging with current facilities, but it would be within reach of the ELTs. In addition to reflected light, this method may also be sensitive to OI auroral emission \citep{luger16}.

In the remainder of this paper, we focus on method (iii).  The basic idea is that a tidally locked planet may have a temperature gradient from the dayside to the nightside. As the planet orbits, the fraction of the dayside that is visible varies.  The thermal phase variation can be predicted exactly for a planet with no atmosphere, assuming the inclination is known via method (ii). If an atmosphere is present, it tends to redistribute the heat to the nightside, thus lowering the amplitude and changing the color of the thermal phase variation. This idea has been discussed before \citep{gaidos04, seager09, selsis11, maurin12, selsis13}, and and applied specifically to climate models of Proxima b by \cite{turbet16}. Here we perform detailed signal-to-noise calculations and simulate a retrieval of the planet's atmospheric properties based on possible observations with the \project{James Webb Space Telescope} (\jwst, scheduled for launch in 2018).

\section{Methods}
\label{sec:methods}
\subsection{Toy Climate Model}
To predict the thermal emission from Proxima b, we require a model of its climate.  We use a simple model that makes the following assumptions: if no atmosphere is present, all the incident stellar flux will be absorbed and reradiated on the planet's dayside as a blackbody. We assume the emissivity of the surface is unity; for more discussion of this point, see \S\;\ref{sec:assumptions}. On the other hand, if there is an atmosphere, it can advect heat to the nightside.  

We also assume the planet is tidally locked, and check this assumption by calculating the tidal locking timescale from \cite{gladman96}. For a tidal $Q$ of 100 and an initial rotation rate of one cycle per day, the locking timescale is $t_\mathrm{lock} \sim 10^4\,\mathrm{yr}$, which is short compared to the age of the system.  However, we note that if the planet has a non-zero eccentricity, it is possible that the locking timescale if significantly longer \citep{coleman16, ribas16}.  Continued radial velocity monitoring of the system will be important for precise constraints on the eccentricity. We also note that moons could potentially delay tidal locking of the planet to the star; however, moons are unlikely to be present around Proxima b due to instability of their orbits on gigayear timescales \citep{sasaki14}.


Based on these assumptions, we use an analytic climate model of the form:\\

\noindent $\sigma T^4 = $
\vspace{-3mm}

\begin{equation*}
\begin{cases}
S \times (1 - A)\times F/2, & \pi/2 < \left|\theta\right| < \pi \\
S \times (1 - A) \times [F/2 + (1 - 2F)\cos{z}], & \left|\theta\right| \le \pi/2
\end{cases}
\end{equation*}
where $S= 0.65\, S_\oplus = 890\,\mathrm{W/m^2}$ is the Proxima b irradiance, $A$ is the Bond albedo, $0 < F < 0.5$ is the fraction of incident solar energy advected to the nightside, $z$ is the zenith angle, $\theta$ is the longitude, and $\sigma$ is the Stefan-Boltzmann constant. We neglect internal heat flux because it contributes a negligible fraction of the total energy budget \citep[assuming an Earth-like heat flux of order $100\; \mathrm{mW/m^2}$][]{davies10}.

\begin{figure}
\resizebox{\hsize}{!}{\includegraphics{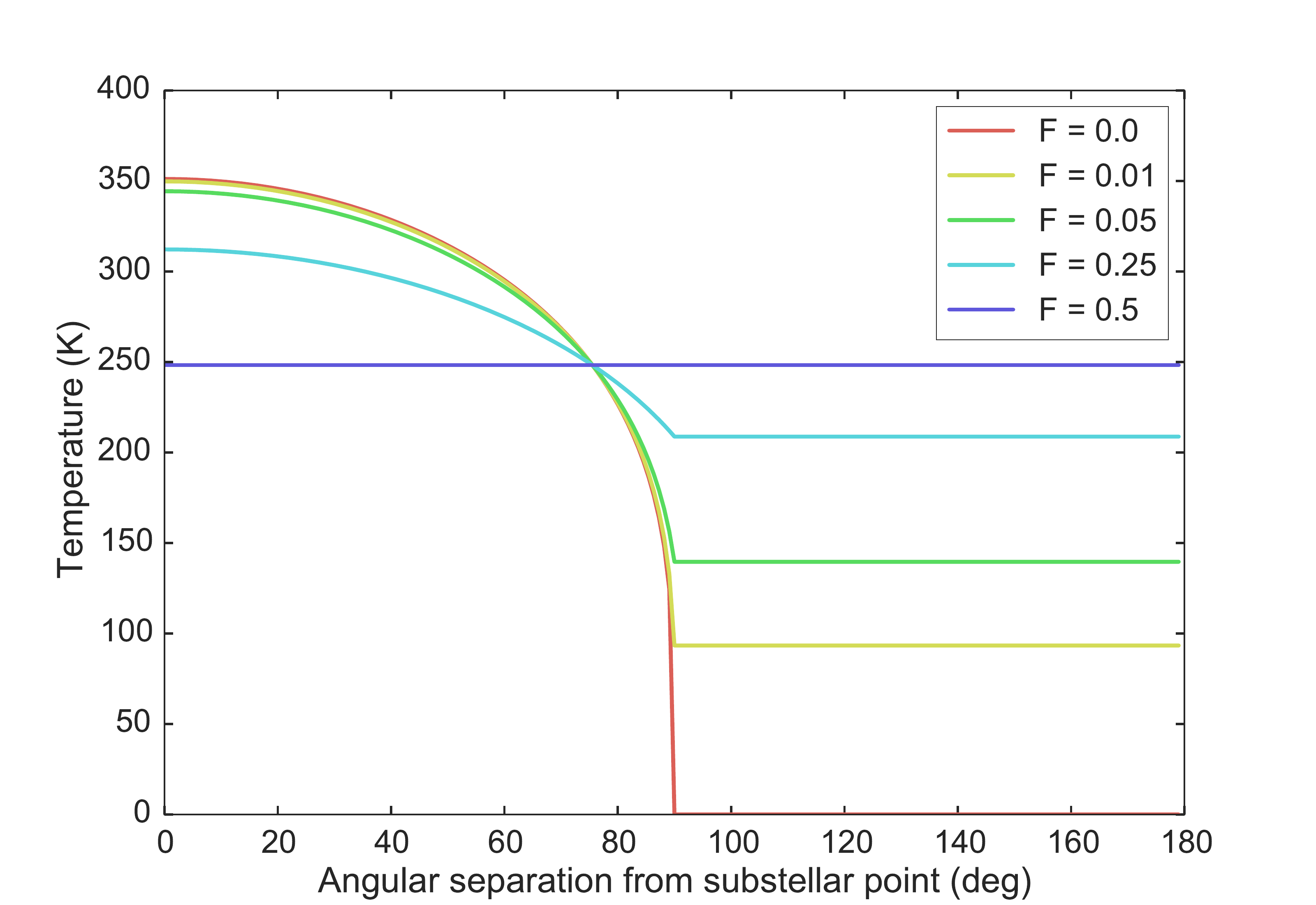}}
\caption{Temperature maps for a range of values for the heat redistribution parameter $F$. This calculation assumes an insolation $S = 890 \;\mathrm{W/m^2}$ and an albedo of 0.1.}
\label{fig:Tmap}
\end{figure}

The above expression behaves well for the limiting cases $F = 0$ and $F = 0.5$.  For the zero redistribution case, the dayside incident flux is proportional to the cosine of the zenith angle, and the nightside recieves zero flux, in agreement for the expected climate of a planet with no atmosphere. For the other limiting case  $F = 0.5$, where half the incident flux is redistributed to the nightside, the planet is isothermal.  In Figure\;\ref{fig:Tmap}, we show example temperature maps for a range of values for $F$.

\begin{figure*}
\resizebox{\hsize}{!}{\includegraphics{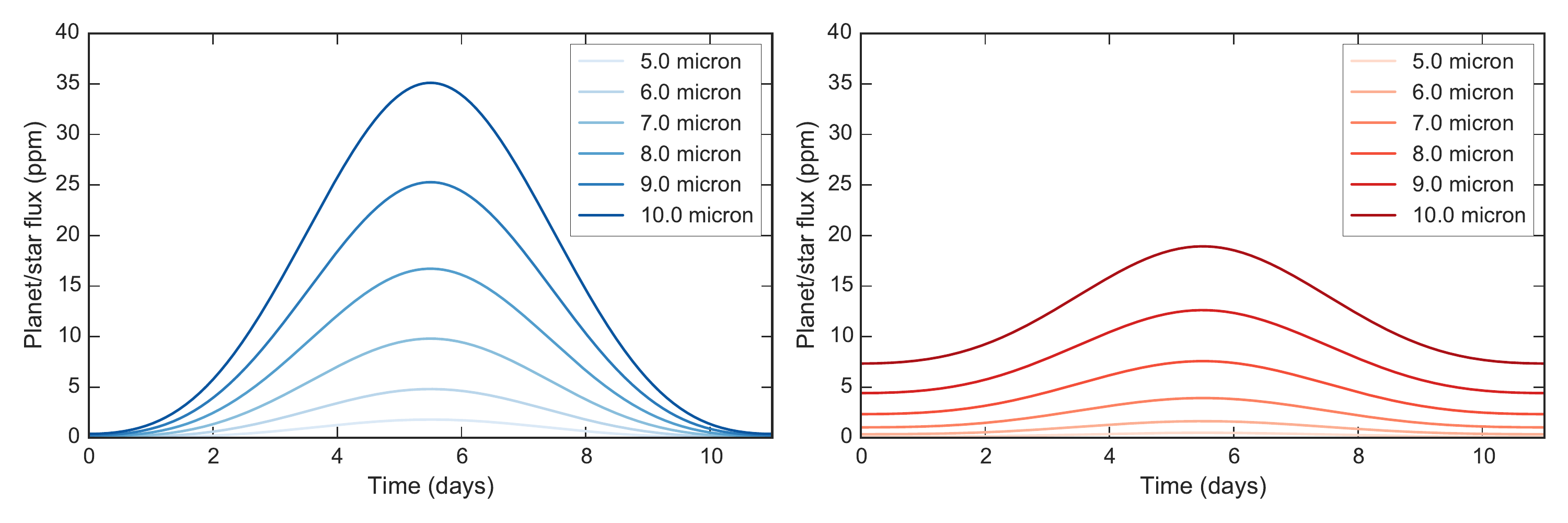}}
\caption{Thermal phase curves for a bare rock (left) and a planet with 35\% heat redistribution. The models both assume an inclination of 60 degrees and an albedo of 0.1.}
\label{fig:phasevar}
\end{figure*}

\begin{figure}
\resizebox{\hsize}{!}{\includegraphics{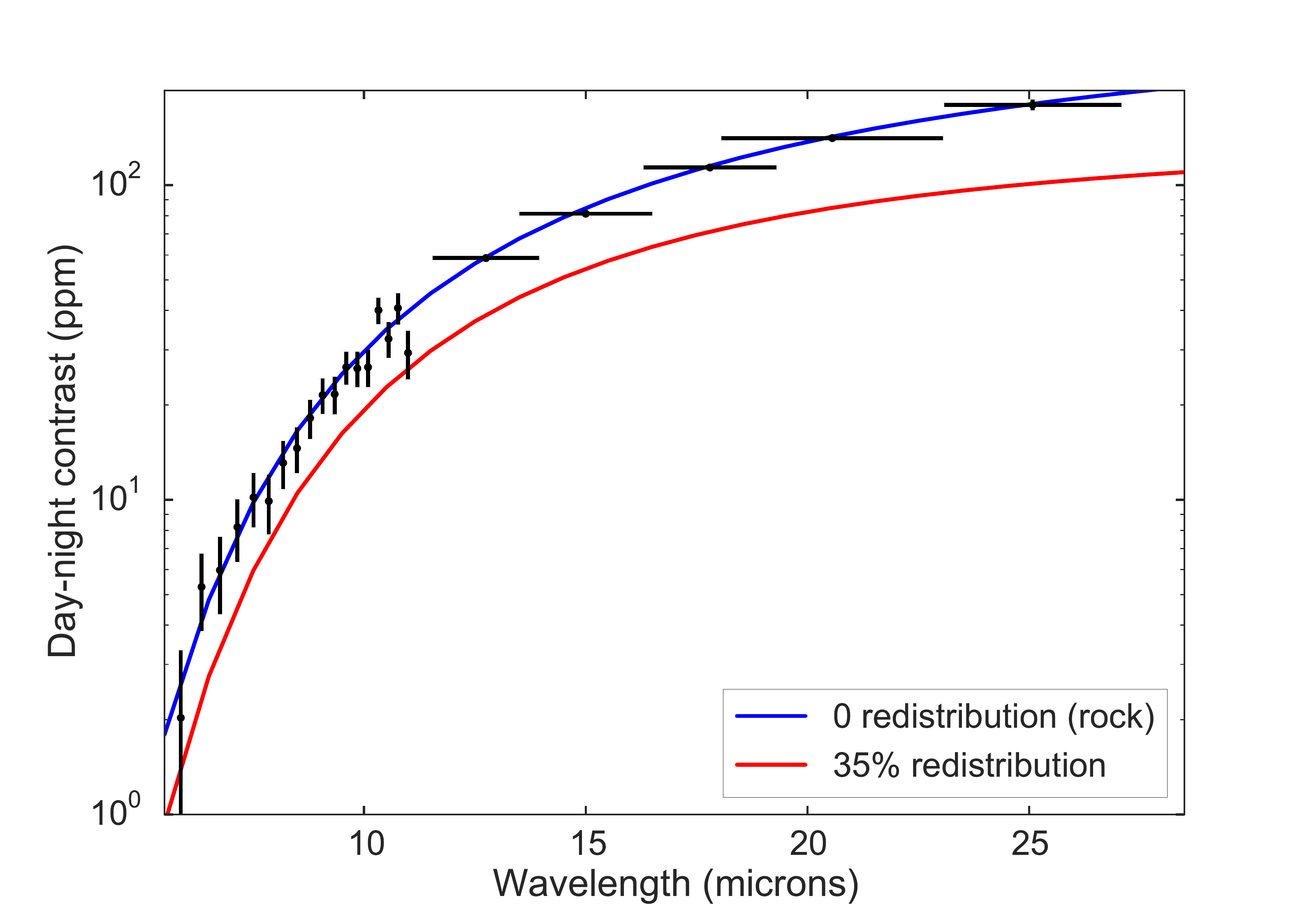}}
\caption{The phase variation spectrum for Proxima b. The models (blue and red curves) correspond to the difference between the measured star+planet spectrum at phase 0.5 and at phase 0.0 for the case of a rock (no heat redistribution, blue), and a planet with an atmosphere that advects 35\% of the heat to the nightside (red). The data points are simulated MIRS/LRS measurements from 5 - 12 $\mu$m and MIRI imaging measurements $>12\,\mu$m. The uncertainties for the simulated data are based on the photon noise for the difference between two phase curves bins, where the spectrum in each bin is co-added over an integration time of 24 hours.} 
\label{fig:spectrum}
\end{figure}

\subsection{\jwst/MIRI signal-to-noise calculation}
To calculate the feasibility of detecting Proxima b's thermal emission with \jwst\, we used the beta version of the \jwst\ Exposure Time Calculator (ETC, available at \url{jwst.etc.stsci.edu}) to estimate signal-to-noise (SNR) for MIRI observations of Proxima Cen.  We considered observations with the LRS spectrograph using the slitless mode optimized for exoplanet observations \citep{kendrew15} as well as photometric imaging observations at $\lambda > 12 \mu$m. We did not consider MRS spectroscopy, because MRS is an integral field spectrograph, and slit losses are a major concern for precision exoplanet atmosphere characterization \citep{beichman14}. 

For our model spectrum, we assumed a 3000 K blackbody normalized to the K-band magnitude of Proxima Centauri. Line blanketing in the optical and near-IR can cause the spectrum to depart from a blackbody, but this effect is weaker at the wavelengths sensed by MIRI where fewer lines are present. To check whether line blanketing affects the normalization of the spectrum, we compared a PHOENIX stellar atmosphere model to a blackbody and found that at K-band they agree to within 10 percent \citep{husser13}.

From the input stellar spectrum, the ETC produces the expected count rate per resolution element in photoelectrons per second. For slitless LRS spectroscopy, which has a resolution of 100 at $7.5\,\mu$m, the count rate ranges from $3.2\times 10^6\;\mathrm{e/s/resolution\;element}$ near $5.5 \, \mu$m to $4\times10^{5}$ at $10\,\mu$m. For the filters, central wavelengths of $\lambda = [12.8, 15, 17.8, 20.6, 25.1]$ have count rates of $[1.3\times10^7, 1.0\times10^7, 4.1\times10^6, 5.8\times10^5]$. These values are in good agreement with signal-to-noise predictions from \cite{cowan15}.

For LRS spectroscopy, the Proxima spectrum saturates the detector over the range $5 - 6\,\mu$m, even in the shortest exposure time (0.15 sec).  However, there is a possibility of implementing alternate readout modes (Nikole Lewis, priv. comm.) that could decrease the number of saturated pixels and also improve the duty cycle of the observations.

In either case, our science goals are not strongly affected by saturating the bluest pixels ($<6\;\mu$m) because the signal is small (a few ppm) at those wavelengths.  The photometric observations do not saturate as quickly and can therefore achieve 80\% or higher duty cycle.  For our final SNR calculations, we assumed a 50\% duty cycle for LRS and an 80\% duty cycle for the photometry. We also assumed that the noise is photon-limited; i.e., uncertainty due to background and flat-fielding are negligible.  

\section{Results}
\subsection{Predicted thermal phase variation}
We used the toy climate model to predict the IR phase variation of Proxima b over the course of its orbit. We considered two scenarios: a ``rock" case, with zero assumed heat redistribution, and an ``atmosphere" case, with  moderate redistribution ($F = 0.35$). 

The atmosphere case is motivated by sophisticated GCM modeling of the atmospheric circulation patterns for tidally locked terrestrial planets \citep[e.g.][]{joshi97, merlis10, heng11a, heng11b, pierrehumbert11, selsis11, leconte13, yang13, yang14, koll15, koll16, turbet16}.  These studies have shown that the presence of an atmosphere can reduce the amplitude of infrared phase variation by a factor of two or more.  We therefore tuned the redistribution parameter so that the phase amplitude is half that of a rock at 10 $\mu$m.

For both scenarios, we assumed an inclination of $60^\circ$ (the median value for an isotropic distribution of inclinations) and a Bond albedo of 0.1 \citep[a typical value for rocky bodies;][]{usui13}. Physically realistic atmospheres would likely have a higher albedo (e.g., Earth's albedo is 0.3). However, assuming the same albedo is a more conservative choice because it means the planet's properties are harder to distinguish.  We used the values reported in \cite{anglada16} for the planet's physical and orbital parameters. We assumed a planet radius of $1.1 \, R_\oplus$, based on predictions from the terrestrial planet mass-radius relation \citep{chen16}. See \S\;\ref{sec:assumptions} for a discussion of the uncertainty in the planet-to-star radius ratio. 

Figure\,\ref{fig:phasevar} shows the predicted thermal phase curves in the wavelength range $5-10\;\mu$m for the two cases. We calculate the phase curve directly from the temperature map using the \texttt{SPIDERMAN} software package for Python \citep[in development on GitHub at \url{https://github.com/tomlouden/SPIDERMAN}][]{louden16} . The peak-to-trough phase variation at 10 $\mu$m for the rock case is 35 ppm. The amplitude is sensitive to wavelength, varying by over an order of magnitude between 5 and 10 $\mu$m. This strong wavelength dependence results from the ratio of blackbody intensities for the planet versus the star. The planet's emission peaks near 10 $\mu$m, whereas the star peaks in the optical and decreases steeply with wavelength. The atmosphere case ($F$ = 0.35) shows a similar wavelength dependence in the phase curve amplitude; however, the overall amplitude is scaled down by a factor of two compared to the rock.

\subsection{Simulated Spectrum and Retrieval of Atmospheric Properties}
Using the climate model and \jwst\ noise estimates described in \S\;\ref{sec:methods}, we simulated a measurement of the thermal phase variation for Proxima b. Following \cite{selsis11}, we define the phase variation as the difference between the star + planet spectrum at phase 0.5 and phase 0.0. We show the results in Figure\;\ref{fig:spectrum}. We plot simulated data for LRS as well as all the photometric filters. However, note that each data set (LRS + each filter individually) requires a complete phase curve observation to obtain.  Full phase coverage is required because the detectors are expected to have percent-level sensitivity variations over time, which make it impossible to stitch together segments of the phase curve observed at different epochs. 

We wish to know how robustly we can determine the heat redistribution on the planet based on these measurements. The key parameters that the spectrum depends on are the orbital inclination, the planet-to-star radius ratio, the albedo, and the heat redistribution. We assume the inclination is known exactly and that the radius ratio is known to a precision of 10\% (for discussion of this point, see section \S\;\ref{sec:assumptions}).  The remaining unknowns are thus the albedo $A$ and the heat redistribution $F$.

\begin{figure}
\resizebox{\hsize}{!}{\includegraphics{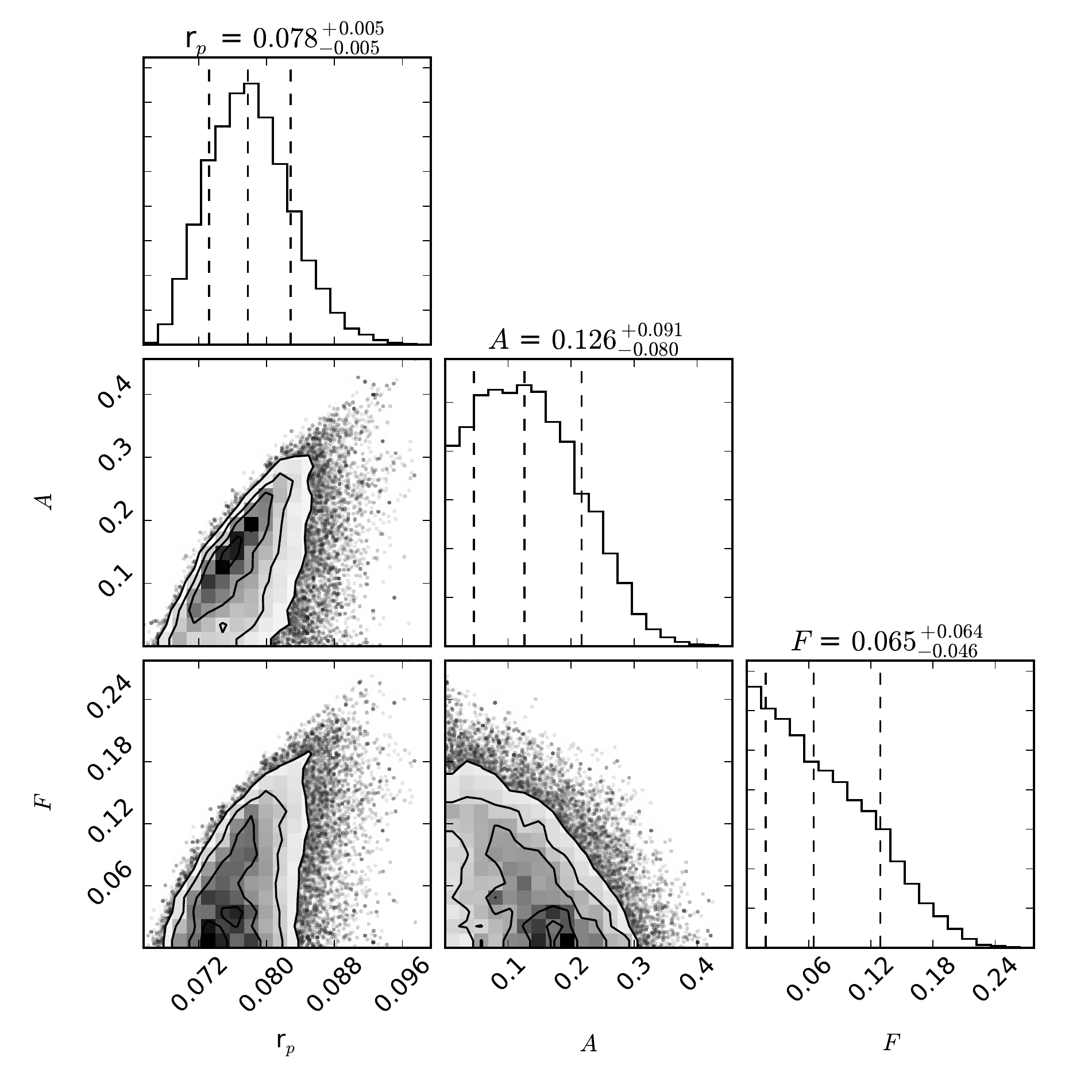}}
\caption {Pairs plot for MCMC fit to the MIRI/LRS thermal variation spectrum showing the posterior distributions for the planet-to-star radius ratio $r_p$, the Bond albedo $A$, and the fractional heat redistribution $F$. }
\label{fig:pairs}
\end{figure}

To assess how tightly we can constrain the albedo and heat redistribution,  we ran an MCMC fit to the simulated LRS spectrum from Figure\,\ref{fig:spectrum}, assuming a fixed inclination, a Gaussian prior on the planet radius with standard deviation 10\% of the best fit radius value, and albedo $A$ and redistribution $F$ were free parameters.  We used the \texttt{emcee} package to perform the fit \citep{fm13}. Figure\;\ref{fig:pairs} shows the resulting posterior distribution of the fit parameters. We measure the heat redistribution to be $F = 0.07^{+0.06}_{-0.05}$, which is inconsistent with the moderate redistribution atmosphere case at 4.5\,$\sigma$ confidence. The albedo is also well constrained, to $A = 0.13^{+0.09}_{-0.08}$. These results demonstrate that a single MIRI/LRS phase curve observation is a powerful diagnostic of the presence of an atmosphere on Proxima b.

\subsection{The 10\,$\mu$m ozone feature}
\label{subsec:ozone}
We also explored the feasibility of detecting an ozone absorption feature from the planet at 10 $\mu$m. This is a prominent feature of Earth's IR emission spectrum, and noteworthy as a potential biosignature \citep{segura05,lin14}, though it can also arise from arise abiotically from evaporated oceans \citep[e.g.][]{ribas16}. For this case, we assumed Earth-like atmospheric properties: a Bond albedo $A = 0.3$ and an isothermal temperature structure.  We then scaled the planet's blackbody signal by a model for the fractional emergent IR spectrum calculated by \cite{rugheimer15}. This model assumes an Earth-like atmospheric composition irradiated by a GJ 1214b-like star. We note that the presence of ozone is sensitive to the ultraviolet (UV) spectrum of the host star \citep{rugheimer15}, which can vary for M-dwarfs even of the same spectral type \citep{france16}. UV spectroscopy of Proxima should therefore be a priority while it is still possible with the \project{Hubble Space Telescope}.

\begin{figure}
\resizebox{\hsize}{!}{\includegraphics{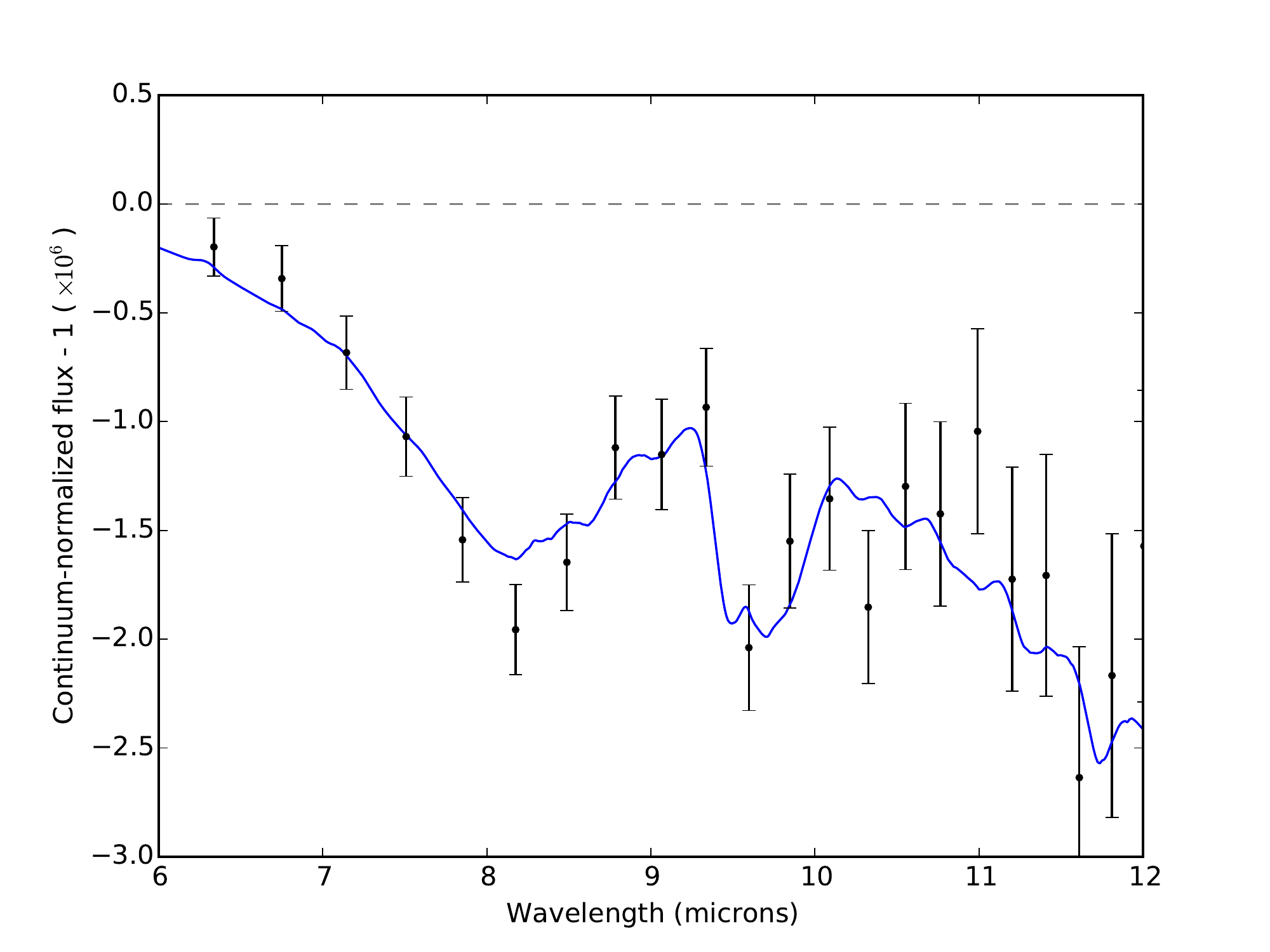}}
\caption{Continuum normalized star + planet spectrum (blue line). The absorption feature centered at 9.8 $\mu$m corresonds to an ozone band. The feature at 8 $\mu$m is due to methane, which is also unexpected in the stellar spectrum \citep[assuming equilibrium chemistry;][]{heng16}. The simulated data assumed photon-limited precision from 60 days of co-added observations.}
\label{fig:ozone}
\end{figure}

The continuum normalized spectrum of the star + planet is shown in Figure\;\ref{fig:ozone}. The ozone feature does not vary with planet orbital phase, but it is in principle detectable from a very high signal-to-noise combined spectrum , because M-dwarfs are too hot to have abundant ozone in their photospheres. However, the predicted feature amplitude is small -- less than one part per million over a narrow band.  For a qualitative illustration of how much observing time is required to detect the feature, we plotted a simulated spectrum co-added from 60 days total integration.  We note that this model spectrum is the most challenging case to detect. If there is a temperature contrast between the day and nightside, the ozone feature depth on the dayside would be a factor of several larger than for the isothermal scenario, and in addition, the periodicity of the signal would help distinguish it from variation in the stellar continuum due to changing star spot coverage. In any case, such an observation would be tremendously exciting if successful, but we discuss several important caveats regarding feasibility in the next section. 

\section{Assumptions \& Caveats}
\label{sec:assumptions}
In our analysis, we make several important assumptions about the planetary system and the data obtainable with \jwst, which we outline below:
\begin{enumerate}
\item{\emph{The star is a perfect blackbody.} In reality, the stellar spectrum has a forest of atomic and molecular absorption lines (mainly due to water). Model infrared spectra for mid-M dwarfs depart from a blackbody at the $1\%$ level at the wavelength and resolution of MIRI/LRS \citep{veyette16}, and absorption features will change in amplitude as star spots with varying water content rotate in and out of view. Proxima's star spot properties not known, but assuming $1\%$ variability (appropriate for a 1\% covering fraction and 300 Kelvin temperature difference), the stellar spectrum will vary at the 100-ppm level. Correcting for this effect is particularly important for the detection of the 9.8 $\mu$m ozone feature, which is only one ppm in amplitude. Even for the thermal phase variation, which is periodic and larger in amplitude, the changing starspot coverage could pose a significant challenge. The star's rotation period is 83 days \citep{anglada16}, but individual spots can rotate out of view on timescales comparable to the planet's orbit.  Therefore, robust detection of the planet signal will require improved stellar models and water lines lists in the infrared \citep{fortney16}, as well as detailed characterization of the stellar spot coverage. Before undertaking an intensive \jwst\ observing campaign, it will be important to assess whether these improvements are feasible at the level of precision required.} 
\item{\emph{The precision of the measurements is photon-limited.} Past observations with space-based telescopes have been successful at reaching the photon limit \citep{kreidberg14, ingalls16}.  These results are encouraging; however, they have not approached ppm-level precision, and MIRI has a different type of detector (arsenic-doped silicon). Testing the precision of the MIRI detectors early in the mission will be key for guiding potential observations of Proxima Centauri.}
\item{\emph{The inclination and planet-to-star radius ratio can be determined.} These quantities are necessary for interpreting the thermal phase variation.  We assume the inclination will be measured with the combination of high-contrast imaging and high-resolution spectroscopy (method (ii) of \S\;\ref{sec:possibilities}). Past detections of non-transiting planets have determined the inclination to a precision of about $1^\circ$ \citep{brogi12}. The planet-to-star radius ratio can also be estimated precisely:  the mass-radius relation for terrestrial bodies is tight \citep[with scatter less than five percent;][]{dressing15a,chen16}. Therefore, the dominant sources of uncertainty for the planet-to-star radius ratio are the stellar radius, which is known to about 5 percent from interferometric observations \citep{demory09} and the planet minimum mass, which is already known to 10 percent \citep{anglada16}.}
\item{\emph{Heated rock radiates with a blackbody spectrum.} For this to be the case, the emissivity of the rock must be unity. Rocky material tends to have high emissivity in the IR (near 0.9), but the exact value depends on wavelength and the composition of the rock, and can drop as low as 0.5 \citep{karr13}. Detailed modeling of the impact of emissivity on the predicted thermal phase variation is beyond the scope of this paper, but it should be considered in future study of Proxima b.}
\end{enumerate}

\section{Conclusion}
In this paper, we outlined an observational test of the existence of an atmosphere on Proxima b.  By combining intensive observing programs from the ground and space, it is possible to precisely measure the fraction of incident flux that is redistributed to the nightside of the planet. 

In the case of no redistribution, one could infer the planet does not have an atmosphere and is unlikely to host life. By contrast, if we do find evidence for significant energy transport, this would indicate that an atmosphere  or ocean are present on the planet to help transport the energy. In that case, Proxima b would be a much more intriguing candidate for habitability. Either way, these observations will provide a major advance in our understanding of terrestrial worlds beyond the Solar System.

\acknowledgments
We thank Nikole Lewis, Natasha Batalha, and Klaus Pontoppidan for tips about \jwst\ SNR calculations. We are also grateful to Jayne Birkby and Mercedes L\'opez-Morales for insightful discussions about what observations are possible for Proxima Cen b. We appreciate thoughtful comments on the manuscript from Ed Turner and Tom Greene. We thank Colin Goldblatt for graciously sharing his knowledge of rocks. We also thank Sarah Rugheimer and Mark Veyette for making their models publicly available. Finally, we thank the referee for his or her comments, which improved the quality of the discussion.

\bibliographystyle{apj}
\bibliography{ms.bib}

\end{document}